\documentclass[reprint, aps, prl,onecolumn,notitlepage]{revtex4-1}
\usepackage{amsmath}
\usepackage{amsfonts}
\usepackage{amssymb}
\usepackage{graphicx}
\usepackage{fullpage}

\usepackage{mathabx} % for \Asterisk
\providecommand{\U}[1]{\protect\rule{.1in}{.1in}}

\newcommand{\cT}[1]{{#1}^\dagger} % conjugate transpose
\newcommand{\Zof}[1]{\mathcal{Z}\left(#1\right)}
\newcommand{\QQ}{\mathbb{Q}}
\newcommand{\RR}{\mathbb{R}}
\newcommand{\CC}{\mathbb{C}}
\newcommand{\HH}{\mathbb{H}}
\newcommand{\FF}{\mathbb{F}}
\newcommand{\nonz}{\bigast}

\begin{document}

\title{Locality for quantum systems on graphs depends on the number field}
\author{H. Tracy Hall}
\affiliation{Department of Mathematics, Brigham Young University,
Provo Utah 84602, USA.}
\email{h.tracy@gmail.com}
\author{Simone Severini}
\affiliation{Department of Computer Science, and Department of
Physics \& Astronomy, University College London, London WC1E 6BT, UK.}
\email{simoseve@gmail.com}

\begin{abstract}
Adapting a definition of Aaronson and Ambainis [\emph{Theory Comput.} \textbf{1} (2005),
47--79], we call a quantum dynamics on a digraph
\emph{saturated $Z$-local} if the
nonzero transition amplitudes specifying the unitary evolution are in exact
correspondence with the directed edges (including loops) of the digraph.
This idea appears recurrently in a variety
of contexts including angular momentum, quantum chaos, and combinatorial
matrix theory.
Complete characterization of the digraph properties that allow such
a process to exist is a long-standing open question that
can also be formulated in terms of minimum rank problems.
We prove that saturated $Z$-local dynamics involving complex
amplitudes occur on a proper superset of the digraphs
that allow restriction to the real numbers or,
even further, the rationals.
Consequently, among
these fields, complex numbers guarantee the largest possible choice of
topologies supporting a discrete quantum evolution. A similar construction
separates complex numbers from the skew field of quaternions. The result proposes a concrete ground for
distinguishing between complex and quaternionic quantum mechanics.
\end{abstract}


%\keywords{complex numbers, quantum dynamics, quantum information}

\maketitle

\section{Introduction}

Since the early 2000s substantial attention has been given to various types of
quantum dynamics defined with respect to an underlying network, or graph.
Studies have proposed such dynamics, mainly in the context of quantum
information processing, to efficiently induce
useful probability distributions. Relevant applications range from techniques
for searching and exploring combinatorial structures \cite{aa05, cg04} to
settings for universal computation \cite{c08}.
Such models also play a
central role in protocols for state preparation and
transfer in nanodevices based
on spin systems \cite{b03} and in modeling the transport of energy in
biochemical structures \cite{m08}.
In most of these settings a graph specifies a symmetric network of interactions, but
we need to distinguish directions and will work
instead with digraphs.

A digraph is denoted by $G=(V,E)$ where $V$ consists
of $n$ vertices and $E$ is a set of ordered pairs of vertices,
called directed edges.
The source and target of an edge
need not be distinct; we allow loops.
(Equivalently, a digraph is a mathematical relation
on a finite set.)
A canonical way to associate a Hilbert space $\mathcal{H}_{V}$ to $G$ is to define
\( \mathcal{H}_{V}\cong\mathbb{C}^{n}=\left<
|i\rangle:i\in V\right>$,
where $|i\rangle$ is, as usual, the $i$-th element
of the standard basis. The space $\mathcal{H}_{V}$ is the state space of a
scalar quantum particle constrained to evolve on the vertices of $G$.
To each digraph $G$ is associated an adjacency matrix, which, over the class
of digraphs we consider, may be any square matrix with entries in $\{0, 1\}$.
For any vertex with a loop, the corresponding diagonal entry is $1$.
To any matrix combinatorial matrix theory associates a {\em zero-pattern} \cite{b93},
a matrix with entries in $\{0, \nonz\}$ where $\nonz$ stands for an unknown
entry that can take any value other than zero.
(The term ``sparsity pattern'' is also used.)
We adapt terminology
introduced by Aaronson and Ambainis \cite{aa05} (see also Osborne \cite{o08})
to say that a unitary matrix $U\in U\left(  n\right) $ (not necessarily
Hermitian), or the quantum dynamics it specifies, is
{\em saturated $Z$-local} on $G$ if $U$ has the same zero-pattern
as the adjacency matrix of $G$.
Supposing that $H$ is a simple (non-directed) graph and
$F$ is the digraph on the same vertex set
that has a symmetric pair of directed edges
for each edge of $H$ and that has a loop at each vertex,
then a quantum dynamics on $H$ is {\em $Z$-local}
in the sense of \cite{aa05} if
it is saturated $Z$-local on a subdigraph of $F$.

The notion of $Z$-locality is applied to a discrete dynamics,
defined for example by a quantum circuit model,
just as the notion called
$H$-locality applies to evolution under a Hamiltonian. In fact, $H$-locality
expresses locality by the interactions specified by the Hamiltonian operator
(\emph{i.e.}, the clustering of correlations; see \cite{o08}). Aspects of
$Z$-locality for matrices in various stochastic ensembles have been
investigated within the analysis of models describing the time evolution in
quantum graphs and the quantum mechanics of systems that are classically
chaotic \cite{hsw07, ks97, p01, t01}. Recall that a matrix $M$ is \emph{unistochastic
}(resp. \emph{orthostochastic}) if there is a unitary (resp. real orthogonal)
matrix $U$ such that $M_{i,j}=|[u]_{i,j}|^{2}$, for every $i$ and $j$. Because
of this definition, it appears obvious that $Z$-locality can be studied
equivalently in the two settings of stochastic and quantum processes.

Among other areas \cite{dz09}, these matrices occur in foundational questions
\cite{l60, l97} and in high energy physics, where their role is to
characterize interactions between elementary particles,
with the Cabibbo-Kobayashi-Maskawa matrix
\cite{ckm} as the arguably most famous example.

Differences between stochastic ensembles have been discussed since the 1960s
\cite{m63} and it is well-known that the set of unistochastic matrices
includes properly the set of the orthostochastic ones. A complete
characterization of saturated
$Z$-locality is not currently known, although the matrix
analysis literature contains several graph-theoretic and linear algebra
conditions attempting to tackle this problem (see \cite{d11} and the list of
references contained therein). Even if the original question was
mathematically formulated in 1988 \cite{f88}, the connection with minimum rank
problems was underlined only recently \cite{jm08}. The connection is important
because it locates the problem within the context of rank-related topological
invariants including, for example, the Colin de Verdi\`{e}re number
and the Lov\'{a}sz $\vartheta $-function \cite{fh07}.

While it is immediate to see that the zero-patterns of unitary (resp. real
orthogonal) and unistochastic (resp. orthostochastic) matrices are identical,
it is not obvious whether every unitary matrix can
be ``flattened'' to an orthogonal matrix having the same zero-pattern.
As a part of the effort to characterize the ortho- and unistochastic
property \cite{z01}, this question was resolved in the positive for matrices of dimension
$n\leq5$ \cite{d09}, meaning that for small matrices the sets of orthogonal and
of unitary zero-patterns coincide.
It turns out that the general situation is different, as it is shown in the present
work.

For dynamics with amplitudes in
a skew field $\FF$, we study in each case the set
$\Zof{\FF}$ of zero-patterns of
digraphs that allow for saturated $Z$-local dynamics.
We are interested specifically
in the skew fields $\QQ$, $\RR$, $\CC$, and $\HH$ of the rationals,
the reals, the complex numbers, and the quaternions,
respectively, whose sets of $\FF$-orthogonal matrices
(preserving an abstract inner product over $\FF$)
are, respectively, the rational orthogonal, orthogonal, unitary, and hyperunitary
matrices.
The set $\Zof{\FF}$ consists of all
zero-patterns of matrices that are $\FF$-orthogonal.
We shall give a proof of the following statement:

\bigskip

\noindent\textbf{Theorem. }\[\Zof{\QQ} \subsetneq \Zof{\RR} \subsetneq \Zof{\CC} \subsetneq \Zof{\HH}.\]

The proof is constructive and can be used to produce concrete examples of zero patterns belonging to one set but not the other, with dimensions as follows:

\begin{itemize}

\item $\Zof{\QQ} \subsetneq \Zof{\RR}$: $n=35$

\item $\Zof{\RR} \subsetneq \Zof{\CC}$: $n=47$

\item $\Zof{\CC} \subsetneq \Zof{\HH}$: $n=141$

\end{itemize}

The minimum dimensions of separation remain an open question.
\bigskip

A quantum dynamics on a graph $G$ governed by a $Z$-local unitary matrix can
be seen as a generalization of a (discrete)\ random walk on the vertices of
$G$ where the transition probabilities, which define a stochastic matrix, are
substituted by transition amplitudes. Sinkhorn's theorem \cite{mo67} states
that every stochastic matrix can be made bistochastic without modifying its
zero-pattern (recall that a nonnegative real matrix is \emph{stochastic} if
the entries of each row sum up to one; \emph{bistochastic} if also the columns
satisfy this property). However, we know that the zero-patterns of
bistochastic and unistochastic matrices do not coincide \cite{b93}. This means
that not every digraph $G$ supporting a saturated random walk
has a unitary matrix that is saturated $Z$-local on $G$.

We remark that although we have defined $\Zof{\FF}$ over the
class of digraphs, the results of the theorem apply equally well to
the more usual setting of symmetric networks.
Each digraph $G$ on $n$ vertices, with adjacency matrix $A$,
is associated uniquely to a bipartite simple graph $B(G)$ on
$2n$ vertices whose adjacency matrix has a block form
with zero matrices in the diagonal blocks and $A$ and its transpose
in the off-diagonal blocks.
The graph $B(G)$ supports saturated $Z$-local dynamics
over $\FF$ if and only if $G \in \Zof{\FF}$.
In addition, the dynamics of $B(G)$
can without loss of generality be taken to be Hermitian.

The theorem asserts that a characterization of
potentially orthogonal digraphs
must take the field into account.
A corollary of the result is that there are digraphs, and indeed graphs,
that require complex numbers to define a saturated $Z$-local quantum dynamics.
In other words, complex numbers define saturated
discrete quantum walks on a larger set of graphs
than permitted by the reals; the same relationship holds between real and
rational numbers. This last point solves the main open problem formulated in
\cite{d09}.

On the other hand, techniques
have been proposed to define a quantum dynamics on \emph{every} graph
by the aid of some extra degrees of freedom that permit to enforce $Z$-locality on a
topologically equivalent object. The most common techniques of this type are coined quantum
walks, where the dynamics is \emph{lifted} to a product space composed by a
\emph{shift} and a \emph{coin register} \cite{a01} (or, equivalently, with the use of the
graph-theoretic notions of a line digraph \cite{s03}); and Szegedy's
generalization of Markov chains \cite{s04}. In this respect, the consequences of
our theorem are significant when $Z$-locality is not reflected by the use of
\emph{ad hoc} constructions. More generally, $Z$-locality is not an obstacle
from a point of view embracing specific algorithmic applications,
because there are methods to translate between different number fields \cite{g11, mc09}.

The next section gives a proof of our main result after stating a
required lemma (proof in appendix).
We describe in this work a concrete framework to distinguish between complex and quaternionic quantum mechanics. The debate about this topic is briefly addressed in the final section.

\newcommand{\Ename}{\mathcal{C}}

\section{Proof of the theorem}
The three strict inclusions claimed in the theorem
must arise in each case from a zero-pattern that belongs
to one collection but not to the other,
where the only allowed constraint in $\Zof{\FF}$
is to make each entry zero or nonzero.
Such a coarse degree of control makes
it difficult to engineer an obvious
dependence on the numerical field.
We make use of a technique
that allows us to impose an additional special
type of constraint on a small zero-pattern
with the guarantee that the special constraints are
faithfully reflected by a pure zero-pattern
that may be several times larger.

Given a rectangular zero-pattern $T$, we define
a {\em four-way democracy} on $T$ to specify, within
a single column of $T$, four $\nonz$-entries that are required
to have the same magnitude.
(The choice to define four-way democracies
rather than another size
is tied to a particular $7 \times 7$ zero-pattern
whose rigidity properties are
central to the proof of the lemma.)

A pair $(T, \Ename)$, where $T$ is a zero-pattern
and $\Ename$ is a set of four-way democracies on $T$,
is called a {\em constrained zero-pattern}.
Given a constrained zero-pattern $(T, \Ename)$
and a skew field $\FF$,
if a matrix $A$ over $\FF$ has zero-pattern $T$
and has orthogonal columns we call $A$
an {\em orthogonal representation of $T$
over $\FF$.}
If in addition the magnitudes
of entries of $A$ respect all the
constraints of $\Ename$, we say that $A$ is
a {\em constrained orthogonal representation
of $(T, \Ename)$ over $\FF$.}

\bigskip
\newcommand{\lemmify}[1]{{#1}^{\prime}}
\newcommand{\That}{\lemmify{T}}
\newcommand{\phs}[1]{{\bigast_{#1}}}

\noindent\textbf{Lemma. }
Let $(T, \Ename)$ be a constrained zero-pattern.
Then there exists a zero-pattern $\That$, containing $T$ as a
submatrix (so that $(\That, \Ename)$ is also well-defined
as a constrained zero-pattern),
such that for any $\FF \in \{\QQ, \RR, \CC, \HH \}$
the following are satisfied:
\begin{enumerate}
\item
Every orthogonal representation of $\That$ over $\FF$
is also a constrained orthogonal representation of $(\That, \Ename)$ over $\FF$.
\item
Every constrained orthogonal representation
of $(T, \Ename)$ over $\FF$
can be completed to an
orthogonal representation of $\That$ over $\FF$.
\item If the constraints of $\Ename$ affect only a
single column $c$ of $T$, then column $c$ in $\That$
has no $\nonz$-entries other than
the $\nonz$-entries of $c$ in $T$.
\end{enumerate}

\bigskip

Letting $(\mathbb E, \FF)$ represent any consecutive pair
in the sequence $(\QQ, \RR, \CC, \HH)$,
we exhibit a constrained zero-pattern
$(T_{\FF}, \Ename_{\FF})$ such that the zero-pattern $\That_{\FF}$ promised
by the lemma has an orthogonal representation over the skew field
$\FF$
but no orthogonal representation over the field $\mathbb E$.
To indicate which four $\nonz$-entries belong to a four-way democracy,
we give
them the same subscript (including multiple subscripts if a $\nonz$-entry
belongs to more than one four-way democracy).
With this notation, the three constrained zero-patterns
$(T_{\FF}, \Ename_{\FF})$
to which we will apply the lemma are as follows:
\[
\begin{array}{ccc}
(T_{\RR}, \Ename_{\RR}) = &
(T_{\CC}, \Ename_{\CC}) = &
(T_{\HH}, \Ename_{\HH}) = \\
\ \left[
  \begin{array}{c}
    \phs{0\phantom{,1}} \\
    \phs{0,1} \\
    \phs{0,1} \\
    \phs{0,1} \\
    \phs{1\phantom{,0}}
  \end{array}
\right]
,
\ &\
\ \left[
  \begin{array}{cc}
    \phs{0} & \phs{1} \\
    \phs{0} & \phs{1} \\
    \phs{0} & \phs{1} \\
    \phs{0} & 0 \\
      0  & \phs{1}
  \end{array}
\right]
,
& \
\left[
  \begin{array}{cccc}
    \phs{0} & \phs{1} & \phs{2} & \nonz \\
    \phs{0}& \phs{1} & \phs{2} & \nonz \\
    \phs{0} & \phs{1} & \phs{2} & \nonz \\
    \phs{0} &   0  &   0  & \nonz \\
      0  & \phs{1} &   \nonz  & 0 \\
      0  &   0  & \phs{2} & \nonz
  \end{array}
\right]
.
\end{array}
\]

{\em First inclusion: $\Zof{\QQ} \subsetneq \Zof{\RR}$.}
In the constrained zero-pattern $(T_\RR, \Ename_\RR)$
the five nonzero entries must all have the same magnitude.
We apply the lemma to obtain a zero-pattern $\That_\RR$.
Since the matrix $A = \cT{\left[1 1 1 1 1\right]}$ is a constrained orthogonal
representation of $(T_\RR, \Ename_\RR)$,
it can be completed to an orthogonal representation
$\lemmify{A}$ of
$\That_\RR$.
The orthogonal columns of
$\lemmify{A}$
can be scaled and completed to an orthonormal set of columns of a square matrix
$A^{\prime\prime}$,
whose zero-pattern $P$ belongs to $\Zof{\RR}$.
Now suppose by way of contradiction that $P$ also belongs to $\Zof{\QQ}$,
implying the existence of a rational orthogonal matrix
$B^{\prime\prime}$,
with zero-pattern $P$.
Then a subset
$\lemmify{B}$
of the columns of
$B^{\prime\prime}$ is
an orthogonal representation of
$\That_\RR$ over $\QQ$ whose columns are unit vectors, which
by the lemma is also a constrained orthogonal representation of
$(\That_\RR, \Ename_\RR)$ over $\QQ$.
Since $\Ename_\RR$ applies to a single column,
there are exactly five nonzero entries in that column of
$\lemmify{B}$,
all with the same magnitude, namely $1/\sqrt{5}$,
which cannot be the magnitude of a rational number.
By contradiction, $P$ does not belong to $\Zof{\QQ}$,
implying $\Zof{\QQ} \subsetneq \Zof{\RR}$.

{\em Second inclusion: $\Zof{\RR} \subsetneq \Zof{\CC}$.}
The argument proceeds as in the proof of the first inclusion:
$(T_\CC, \Ename_\CC)$ has a constrained orthogonal representation
consisting of two columns over
$\CC$ (whose inner product requires three complex numbers
of the same magnitude to sum to zero)
but does not have a constrained orthogonal representation
over $\RR$.
Thus $\That_\CC$ has an orthogonal representation
over $\CC$ which can be scaled and completed to a unitary matrix of zero-pattern
$P \in \Zof{\CC}$, but $P$ cannot belong to $\Zof{\RR}$ because
any orthogonal representation of $\That_\CC$ over $\RR$
would contain, as a submatrix,
a constrained orthogonal representation
of $T_\CC$ over $\RR$. Thus we have
$\Zof{\RR} \subsetneq \Zof{\CC}$.

\newcommand{\vecv}{\mathbf{v}}

{\em Third inclusion: $\Zof{\CC} \subsetneq \Zof{\HH}$.}
For the previous claim we used that fact that three numbers
$x$, $y$, and $z$ of equal magnitude cannot sum
to zero in $\RR$, but can do so in $\CC$. In fact,
if $\omega_1$ and $\omega_2$
are the two non-real third roots of unity in $\CC$,
we can have $x+y+z = 0$
in precisely two ways:
$\{y = \omega_1 x, z = \omega_1 y\}$
or $\{y = \omega_2 x, z = \omega_2 y\}$.
Now we show that $T_\HH$ does not have a constrained
orthogonal representation over $\CC$.
Suppose on the contrary that it did, and for the moment limit attention
to just the first three rows of $T_\HH$, naming the columns of that submatrix
$\vecv_0, \vecv_1, \vecv_2, \vecv_3 \in \CC^3$.
Since $\vecv_1$ is orthogonal to $\vecv_0$, three numbers of equal magnitude
must sum to zero, with a choice to be made of $\omega_1$ or $\omega_2$.
If $\vecv_2$ is orthogonal to $\vecv_0$ using the same choice,
then $\vecv_2$ will be parallel to $\vecv_1$;
otherwise, since $\omega_2 = \omega_1^2$,
$\vecv_1$ and $\vecv_2$ will be orthogonal.
In neither case can the remaining constraints be
satisfied to construct a
constrained orthogonal representation of the full pattern $T_\HH$ over
$\CC$.
Over $\HH$, however, there is a continuous family
of third roots of unity, and $\vecv_1$ and $\vecv_2$
can be chosen so that both are orthogonal to $\vecv_0$
but so that they are neither parallel nor orthogonal to
each other, and so that $\vecv_1$ is orthogonal
to $\vecv_3$ but $\vecv_2$ is not.
This allows a complete construction of a constrained
orthogonal representation of $T_\HH$ over $\HH$.
It follows that some zero-pattern $P$, which completes the columns
of $\That_\HH$ to a square matrix, separates the sets
$\Zof{\CC}\subsetneq \Zof{\HH}$.

\section{Conclusions}

 Several questions of a mathematical nature remain open: What are the smallest examples of digraphs that distinguish between the number fields? What is the computational complexity for determining whether a given digraph admits a
saturated $Z$-local unitary matrix? (In an email to the authors, S. Aaronson observed that the problem is reducible to the existential theory of reals and it is therefore in PSPACE.) Is there a combinatorial way to characterize these different families of digraphs?

 From the physical point of view, our result indicates a novel,
alternative ground to explore the distinctions between real,
complex (CQM), and quaternionic (QQM) quantum mechanics,
a subject of debate whose roots go back to the 1930s and the axiomatization of Birkhoff and Von Neumann (see \cite{ad94, s60}).
In the study of correlations, quantum mechanics is ``sandwiched'' between classical mechanics
and general probabilistic theories
(see \cite{ca10} for a recent treatment of this point in the graph theoretic framework).
When we consider number fields, we seem to face a similar situation: while the choice of number field does not affect the computational power of the theory \cite{g11, mc09}, there are cogent arguments about the inadequacy of real numbers (\emph{e.g.}, parameter counting for bipartite mixed states, continuity of time, the quantum de Finetti theorem, the need of superselection rules, \emph{etc.} \cite{aa04}).

In the other direction, the status of the connection between CQM and QQM is still unresolved.
The combinatorial tool of
saturated $Z$-local dynamics
may contribute to the separation of these theories in terms of multiparty
correlations obtainable with given resources (such as a fixed Hilbert space dimensionality specified according to the physical system under consideration).
It is possible at the experimental level that observable entanglement measures could be
employed to reach a contradiction (in terms coherent with the known theory)
to show that QQM leads to provably unphysical claims or that it has a particular range of applicability.
The spirit of these latter assertions is clearly speculative
and they suggest directions where further work is required.

\emph{Acknowledgments.} We benefited greatly from conversations with Scott Aaronson, Daniel
Burgarth, Louis Deaett, Leslie Hogben, Reimer K\"{u}hn, James
Louck, Bryan Shader, Wojciech Tadej, Michael Young, and Karol \.{Z}yczkowski.
SS is supported by the Royal Society.

\section{Appendix: proof of lemma}
The lemma used in the proof of our main result asserts that one
can translate the constraints of
a collection of four-way democracies
on a small zero-pattern matrix
into the zero-pattern alone
of a larger matrix.
The mapping of constraints
depends on a special property of
the following $7 \times 7$ matrix,
\[
M =
\left[
  \begin{array}{rrrrrrr}
 -1 & 0 & 0 & 1 & 0 & 1 & 1 \\
 1 & -1 & 0 & 0 & 1 & 0 & 1 \\
 1 & 1 & -1 & 0 & 0 & 1 & 0 \\
 0 & 1 & 1 & -1 & 0 & 0 & 1 \\
 1 & 0 & 1 & 1 & -1 & 0 & 0 \\
 0 & 1 & 0 & 1 & 1 & -1 & 0 \\
 0 & 0 & 1 & 0 & 1 & 1 & -1
\end{array}
\right],
\]
and its zero-pattern, which we call $F$.
It will be convenient to index the rows and columns
of $7 \times 7$ matrices
by the sequence $(0, 1, \dots, 6)$
and to stipulate that index arithmetic
is modulo $7$.
We first observe a few facts that are interesting
but not directly essential to the proof of the lemma:
\begin{enumerate}
\item Entry $(i,j)$ of $F$ is $\nonz$ if and only if
$i-j$ is a quadratic residue modulo $7$.
\item The zero-entries of $F$ give incidence relations
 on the points and lines of a Fano plane.
\item
The $\nonz$-entries of $F$
give the negated entries in a multiplication table
of the imaginary part of the octonions.
\item $M+I$ is the adjacency matrix of a tournament graph.
\item The matrix $\frac12M$ is rational orthogonal.
 \end{enumerate}
The last two points are interesting partly because it is
an open problem \cite{lu06}
whether there exists an orthogonal (or indeed unitary)
matrix of size greater than $3 \times 3$ that is
saturated $Z$-local on a tournament graph.
The heuristic of dimension counting
is pessimistic about the existence
of such a matrix:
Since $\mathrm O(n,\RR)$ is a manifold of dimension $n(n-1)/2$
and each zero-entry introduces one constraint,
non-redundant constraints would give,
for the set of orthogonal matrices that
are saturated $Z$-local on any given tournament graph,
a supposed dimension that is less (by $n$)
than zero-dimensional, or in other words a
decidedly empty set.
Observe, however, that if a loop is added to each vertex of
a tournament graph
the expected dimension is now zero, so that one would
generically expect a finite set of rigid solutions.
Such a looped tournament graph is the form that $F$ takes, and
the special property we will establish for $M$ and $F$
is indeed a form of rigidity.

\medskip

\noindent\textbf{Claim. }
Suppose that $A$ is an orthogonal representation
of $F$ over $\FF$, where $\FF \in \{\QQ, \RR, \CC, \HH\}$.
Then there exist $7 \times 7$ diagonal matrices
$U$ and $D$ over $\FF$, with $U$
additionally hyperunitary,
such that $UAD = M$.
In particular, $A$ is a constrained orthogonal representation
of $(F, \Ename)$ over $\FF$, where
$\Ename$ consists of one four-way democracy
on the $\nonz$-entries of each column.

{\em Proof of claim.}
We construct the matrices $D$ and $U$ in stages, $D$ as a product of seven matrices $D = D_0D_1\dots D_6$
and $U$ as a product of six matrices $U = U_6\dots U_2U_1$, where each $D_i$ or $U_i$ may differ from the identity
matrix in the $(i,i)$ entry but nowhere else. We also recursively define a sequence of matrices
$A_0, \dots, A_6$ and $B_0, \dots, B_6$ in a way
that uniquely specifies, for any $A$, the choice of each $D_i$ or $U_i$:
\begin{itemize}
\item We start with $A_0 =A$.
\item For $ i = 0, \dots, 6$, we choose $D_i$ such that $B_i = A_iD_i$ has entry $(i,i)$ equal to $-1$.
\item For $i = 1, \dots, 6$, we choose $U_i$ such that $A_i = U_iB_{i-1}$ has entry $(i, i-1)$ real and positive.
\end{itemize}
The final result of this is $B = B_6 = UAD$, which (like $M$)
has $-1$ along the diagonal,
and which has positive real entries on the subdiagonal.
(Note that the entire process can be completed without leaving the field $\FF$.)
We name the entries of $B$ as follows:
\newcommand{\xv}{x}
\newcommand{\yv}{y}
\newcommand{\zv}{z}
\[
B =
\left[
\begin{array}{ccccccc}
  -1   &   0   &   0   & \zv_3 &   0   & \yv_5 & \xv_6 \\
 \xv_0 &  -1   &   0   &   0   & \zv_4 &   0   & \yv_6 \\
 \yv_0 & \xv_1 &  -1   &   0   &   0   & \zv_5 &   0   \\
   0   & \yv_1 & \xv_2 &  -1   &   0   &   0   & \zv_6 \\
 \zv_0 &   0   & \yv_2 & \xv_3 &  -1   &   0   &   0   \\
   0   & \zv_1 &   0   & \yv_3 & \xv_4 &  -1   &   0   \\
   0   &   0   & \zv_2 &   0   & \yv_4 & \xv_5 &  -1
\end{array}
\right] = UAD
\]
The entries $x_0$ through $x_5$ are by construction real and positive
but $x_6$ might, a priori, have a non-trivial complex or quaternionic phase.
Since $U$ is unitary and both $\cT{A}A$ and $D$ are diagonal,
the product $\cT{D}\cT{A}\cT{U}UAD$ is also diagonal
and the columns of $B$ are mutually orthogonal.
By cyclic symmetry, any equation derived from the orthogonality of
columns of $B$ must remain true when the set of indices is permuted
cyclically.
\newcommand{\xin}{X}
We introduce the shorthand $\xin_i = 1/\overline{x_i}$.
(The conjugation in this definition is only necessary in the case of $x_6$, whose quaternionic phase
is for the moment undetermined.)
The fourteen variables $\{\xv_i, \xin_i\}$ all commute,
even in the case $\FF = \HH$, since at most one related pair is not real.

From the fact that entry $(1,0)$ of $\cT{B}B$ is zero, we conclude that
$\overline{\xv_1}\yv_0 = \xv_0$ and thus, for all $i$,
\[ \yv_i = \xv_i\xin_{i+1}.\]

From the fact that entry $(2,0)$ of $\cT{B}B$ is zero, we conclude that
$\overline{\yv_2}\zv_0 = \yv_0$ and thus, for all $i$,
\[ \zv_i = \xv_i\xin_{i+1}\xin_{i+2}\xv_{i+3} .\]

From the fact that entry $(4,0)$ of $\cT{B}B$ is zero, we conclude that
$\overline{\zv_4}\xv_0 = \zv_0$ and thus, for all $i$,
\[ \xv_i = \xv_i\xin_{i+1}\xin_{i+2}\xv_{i+3}\xin_{i+4}\xv_{i+5}\xv_{i+6}\xin_{i+7}\]
and, for example,
\[\xin_0\xin_1\xin_2\xv_3 \xin_4\xv_5\xv_6 = 1,\]
which in particular shows that $\xv_6$ is after all real and positive,
so that in all cases $\xin_i = \xv_i^{-1}$.
It is convenient to take logarithms of this equation,
together with its cyclic index permutations,
and write the homogeneous system of linear equations
\[
\left[
\begin{array}{rrrrrrr}
  -1   &  -1   &  -1   &   1   &  -1   &   1   &   1   \\
   1   &  -1   &  -1   &  -1   &   1   &  -1   &   1   \\
   1   &   1   &  -1   &  -1   &  -1   &   1   &  -1   \\
  -1   &   1   &   1   &  -1   &  -1   &  -1   &   1   \\
   1   &  -1   &   1   &   1   &  -1   &  -1   &  -1   \\
  -1   &   1   &  -1   &   1   &   1   &  -1   &  -1   \\
  -1   &  -1   &   1   &  -1   &   1   &   1   &  -1
\end{array}
\right]
\left[
\begin{array}{c}
\ln \xv_0 \\
\ln \xv_1 \\
\ln \xv_2 \\
\ln \xv_3 \\
\ln \xv_4 \\
\ln \xv_5 \\
\ln \xv_6
\end{array}
\right]
=
0
\]
which has unique
solution $\{\ln \xv_i = 0\}$.
This implies $UTD = M$ and
completes the proof of the claim.

We have in $F$ an example of a zero-pattern which is
\emph{unitarily rigid}, meaning that there is a unique unitary matrix with
pattern $F$ up to scaling of rows and columns by complex phase.
A matrix analogous to $\frac12M$ can be constructed of size $p \times p$
for other primes $p = 4k-1$ (for $k$ an integer), putting $(1-k)/k$ on
the diagonal and $1/k$ in the off-diagonal positions
where the ordered index difference is a
quadratic residue modulo $p$, with zeros elsewhere.
In the cases $k=1$ and $k=2$ the resulting zero-pattern is unitarily
rigid, and it is a natural question, to which we do not know the answer,
whether this gives a unitarily rigid zero-pattern
for every prime congruent to $3$ modulo $4$.

The rigidity of $F$ is what allows us to construct
large zero-patterns from small constrained zero-patterns.
Consider the case of the constrained zero-pattern $(T, \Ename)$
where $T$ is the $7 \times 7$ pattern $F$
and where $\Ename$ consists of
four-way democracies on any subset of the columns.
In this special case, $\That = T$ already satisfies the conclusions
of the lemma.
In general we will construct a zero-pattern $\That$ that positions
several copies of $F$ as submatrices (one for each element of $\Ename$),
in each case with one column overlapping the submatrix $T$ and the other six
columns outside of it.
We give the name $F_6$ to the $7 \times 6$ zero-pattern
consisting of the last six columns of $F$.
\newcommand{\Tone}{R}
\newcommand{\Ttwo}{S}

\emph{Proof of lemma.}
Let $(T, \Ename)$ be a constrained zero-pattern.
It suffices to prove a special case of the lemma
in which $\Ename$ is a singleton $\{C_0\}$, since the process can be iterated:
If $\Ename = \{C_0, \dots, C_{n-1}\}$
then we first apply the special case of the lemma to $(T, \{C_0\})$
to obtain $\That_1$, then apply the special case of the
lemma to $(\That_1, \{C_1\})$, to obtain $\That_2$,
and so forth until we obtain $\That_n = \That$.
Assuming then that $\Ename = \{C_0\}$, we start with $T$ and append rows or columns
in three stages to obtain $\Tone$,
$\Ttwo$, and finally $\That$.

First, we obtain the zero-pattern $\Tone$ from $T$ by appending
three rows of all-zeros. Call these the {\em control rows.}

Next, we obtain the zero-pattern $\Ttwo$ from $\Tone$
by appending six new columns, the {\em control columns,}
whose zero-pattern will depend on the
six columns of the pattern $F_6$:
The four rows that
correspond to $C_0$ get the pattern of
rows $0$, $1$, $2$, and $4$ of $F_6$,
and the three control rows
get the zero-pattern of the remaining
rows $3$, $5$, and $6$ of $F_6$.
All other rows are zero.
Now we can say some things about orthogonal
representations of $\Ttwo$.
Let the column of $T$ or of $\Ttwo$ that is
constrained by $C_0$ be called $c_0$,
and call the control columns $c_1, \dots, c_6$.
Other than rows containing at most
a single $\nonz$,
the zero-pattern of $c_0$ and the control columns is $F$.
It follows that any orthogonal
representation of $\Ttwo$ over $\FF$
is also a constrained orthogonal representation
of $(\Ttwo, \{C_0\})$ over $\FF$, so that
$\Ttwo$ already fulfills the first conclusion
of the lemma.
However, there may still be constrained orthogonal
representations of $(T, \{C_0\})$ over $\FF$
that cannot be completed to an orthogonal representation
of $\Ttwo$ over $\FF$.
This can happen because
the $\nonz$-entries of some non-$c_0$ column $a$ of $T$
and the $\nonz$-entries of some control column $b$
may overlap in as many as two rows.
The purpose of the third stage is to remedy any resulting non-orthogonality.

Finally, we obtain the zero-pattern $\That$ from $\Ttwo$
by appending several additional rows, in each of which
exactly two entries are $\nonz$ and the rest are zero.
For every pair $(a,b)$ consisting of a column $a \ne c_0$ from the original
zero-pattern $T$ and a column $b$ from the
control columns, we
append either zero, one, or two rows with $\nonz$-entries
at $a$ and $b$.
The \emph{mutual support} of $a$
and $b$ is the set of rows in $\Ttwo$
where column $a$ and column $b$
both take the value $\nonz$; this can either be
empty or can consist of one or two rows.
If the mutual support of $a$ and $b$ consists
of exactly one row, then for the
pair $(a, b)$ we append exactly one new row while building
$\That$.
The two new $\nonz$-entries in this row have no other constraints
and only ever interact with each other, and so
columns $a$ and $b$, which in $\Ttwo$ were
forced to be non-orthogonal, can now easily be
made orthogonal in $\That$.

If the mutual support of $a$ and $b$ is empty,
then columns $a$ and $b$ are automatically
orthogonal, and we append no additional row for the pair $(a, b)$.

If the mutual support of $a$ and $b$ consists of two rows,
then regardless of $\Ttwo$
we may, to ensure that columns $a$ and $b$ can easily
be made orthogonal, always safely append two rows for the pair
$(a, b)$ when building $\That$.  (If there is reason
to prefer a smaller $\That$,
in most cases inspection is sufficient to assure that appending either one
additional row or none at all for the pair $(a, b)$ will serve just as well.)

Each appended row or pair of rows in $\That$
allows for the orthogonality of a pair of columns
$a$ and $b$ without spoiling the orthogonality
previously secured for
any other pair of columns.
The complete process of appending rows ensures that
every constrained orthogonal representation
of $(T, \{C_0\})$ over $\FF$ can be completed
to an orthogonal representation of $\That$
over $\FF$. The conclusion also remains,
as it did for $\Ttwo$, that the zero
pattern $F$ occuring as a submatrix
forces
every orthogonal representation
of $\That$ over $\FF$ to be a constrained
orthogonal representation of $(\That, \{C_0\})$
over $\FF$.
Since column $c_0$ is always distinct from $a$ and $b$,
it acquires no new $\nonz$-entries in this process,
which means, as the process is iterated, that
if all the constraints in $\Ename$ affect only
a single column $c$, then column $c$ in
$\That$ has no $\nonz$-entries other
than the $\nonz$-entries of $c$ in $T$.
This completes the proof of the lemma.

\end{document}